\documentclass[aps,prl,twocolumn,nofootinbib,superscriptaddress]{revtex4}
\usepackage{latexsym}
\usepackage{hyperref}
\usepackage{amssymb}
\usepackage{epsfig,amsmath,graphics}
\usepackage{pstricks}
\usepackage{color}
\usepackage{dcolumn}
\usepackage{ulem}
\usepackage{xcolor}
\usepackage{placeins}
\usepackage{lipsum}
\usepackage{tikz-feynman}[luatex=true]
\usepackage{feynmp-auto}
\usepackage{soul}

\definecolor{linkcolor}{rgb}{0.0, 0.28, 0.67}
\hypersetup{colorlinks=true,
	linkcolor=linkcolor,
	citecolor=linkcolor,
	urlcolor=linkcolor,
	linktocpage=true
}

\newcommand{\Mpl}{M_\text{pl}}

\usepackage{comment}
\newcommand{\be}{\begin{equation}}
	\newcommand{\ee}{\end{equation}}

\def\bea{\begin{eqnarray}}
	\def\eea{\end{eqnarray}}
\def\ltap{\ \raise.3ex\hbox{$<$\kern-.75em\lower1ex\hbox{$\sim$}}\ }
\def\gtap{\ \raise.3ex\hbox{$>$\kern-.75em\lower1ex\hbox{$\sim$}}\ }
\def\lsim{\ \raise.3ex\hbox{$<$\kern-.75em\lower1ex\hbox{$\sim$}}\ }
\def\gsim{\ \raise.3ex\hbox{$>$\kern-.75em\lower1ex\hbox{$\sim$}}\ }

\usepackage{slashed}

\usepackage{color}

\newcommand{\ignore}[1]{}
\newcommand{\beq}{\begin{equation}}
	\newcommand{\eeq}{\end{equation}}
\newcommand{\bear}{\begin{eqnarray}}
	\newcommand{\eear}{\end{eqnarray}}

\def\lcdm{$\Lambda{\rm CDM}$}

\newcommand{\Neff}{N_{\rm eff}}
\newcommand{\NeffBBN}{N_{\rm eff}^{\mathrm{BBN}}}

\newcommand{\TDS}{T_{\rm DS}}

\newcommand{\gammaD}{\gamma_{\scriptscriptstyle d}}
\newcommand{\alphaD}{\alpha_{\scriptscriptstyle d}}

\newcommand{\DS}{\ensuremath{\mathrm{DS}}}
\newcommand{\DR}{\ensuremath{\mathrm{DR}}}
\newcommand{\Tosck}{\ensuremath{T_{\rm osc}^{\rm kin}}}
\newcommand{\Tosc}{\ensuremath{T_{\rm osc}^{\rm std}}}

\newcommand{\mphi}{\ensuremath{m_{\phi}}}
\newcommand{\Ho}{\ensuremath{H_0}}

\begin{document}
	%%%%%%%%%%%%%%%%%%%%%%%%%%%%%%

	\title{Evaporating Axion Dark Matter and the Hubble Constant}
	
	\author{Daniel Aloni}
	%\email{}
	\affiliation{Physics Department, Boston University, Boston, MA 02215, USA}
	\affiliation{Department of Physics, Harvard University, Cambridge, MA 02138, USA}
	\author{Hengameh Bagherian}
	%\email{}
	\affiliation{Department of Physics, Harvard University, Cambridge, MA 02138, USA}
	\author{Rashmish K.~Mishra}
	%\email{}
	\affiliation{Department of Physics, Harvard University, Cambridge, MA 02138, USA}%%%%%%%%%%%%%%%%%%%%%%%%%%
	\begin{abstract}
		Axion-like particles are a well-motivated dark matter candidate that can form a condensate with low momentum and high occupation number. In the presence of dark radiation, this condensate loses energy, naturally increasing the energy density of the universe around matter-radiation equality without requiring additional inputs. This general mechanism may offer a solution to the Hubble tension. 
	\end{abstract}
	%%%%%%%%%%%%%%%%%%%%%%
	
	\pacs{95.35.+d}
	\maketitle
	
	%%%%%%%%%%%%%%%%%%%%%%%%%%%%%%%
	\section{Introduction}\label{sec: Introduction}
	Dark matter (DM) is one of the most mysterious components of our universe, comprising almost all of matter, yet its fundamental properties like mass and interactions remain unknown. In the past decade, another major puzzle has emerged: the Hubble tension—a significant discrepancy between the direct measurements~\cite{Riess:2021jrx, Freedman:2019jwv, Scolnic:2023mrv} of the universe's expansion rate and those inferred from Cosmic Microwave Background (CMB) data~\cite{Planck:2018vyg}, now exceeding 5$\sigma$ significance. As the precision of measurements has improved and systematics are better understood, the need for a theoretical explanation has become increasingly urgent. 
	
	Unlike dark matter, for which there is an abundance of theoretical models consistent with the data, the Hubble tension is yet to be resolved by a compelling model. Many approaches to alleviate the tension suggest reducing the sound horizon $r_s$~\cite{Bernal:2016gxb, Aylor:2018drw, Hou:2011ec}. CMB data connects the sound horizon to the inferred value of the Hubble constant today through the precisely measured angular size of the sound horizon $\theta_s = r_s/D_A$. Here 
	$D_A \propto H_0^{-1}$ 
	is the distance to the surface of last scattering, and the inferred value of \Ho{} must increase to keep $\theta_s$ constant. In particular, $r_s$ can be reduced by increasing the energy density before matter-radiation equality (MRE)~\cite{Brust:2017nmv,Blinov:2020hmc,RoyChoudhury:2020dmd,Brinckmann:2020bcn,Kreisch:2019yzn,Escudero:2019gvw,Agrawal:2019lmo,EscuderoAbenza:2020egd,Escudero:2021rfi,Karwal:2016vyq,Poulin:2018cxd,Lin:2019qug,Smith:2019ihp,Bansal:2021dfh,Cyr-Racine:2021alc,Niedermann:2020dwg, Poulin:2023lkg, Garny:2024ums, Chatrchyan:2024xjj, Buen-Abad:2015ova, Chacko:2016kgg, Aloni:2021eaq, Buen-Abad:2022kgf, Abdalla:2022yfr,Schoneberg:2023rnx,Schoneberg:2022grr,Aloni:2023tff, Gonzalez:2020fdy}. Notably, models like Early Dark Energy and Wess-Zumino Dark Radiation~\cite{Aloni:2021eaq} in which the increase of the energy density is localized around MRE, give a better fit to CMB data compared to other cosmological models.
	
	Among the leading theoretical proposals for dark matter, axions and axion-like particles (ALPs) stand out as promising candidates~\cite{Preskill:1982cy, Dine:1982ah, Abbott:1982af}. These compact scalar fields can have very low masses, arising naturally as Goldstone bosons in many extensions of the Standard Model (SM). The most well-known example is the QCD axion, which solves the strong CP problem~\cite{Peccei:1977ur, Peccei:1977hh, Weinberg:1977ma, Wilczek:1977pj}, but other ALPs, such as those predicted in string theory~\cite{Witten:1984dg, Svrcek:2006yi, Arvanitaki:2009fg, Demirtas:2018akl, Demirtas:2021gsq}, do not necessarily interact with SM particles. Instead, they may belong to a richer dark sector with its own particle content and interactions, resembling the complexity of the SM. For the remainder of this paper, we will focus on ALPs residing in the dark sector that interact with the SM only through gravitational coupling.
	
	As dark matter candidates, ALPs exhibit a rich phenomenology. The ALP field’s expectation value oscillates around the minimum of its potential, leading to macroscopic behavior similar to that of massive cold particles~\cite{Marsh:2015xka}, like WIMPs. However, unlike WIMPs, ALPs can be produced non-thermally, allowing them to have much lower masses than the temperature of the universe while still serving as dark matter.
	
	In this letter, we explore the phenomenology and cosmology of an ALP that feebly interacts with a dark sector composed of dark radiation. This dark radiation consists of massless particles that interact with each other. Specifically, we are interested in a sector of dark radiation that behaves as a perfect fluid. Our main observation is that ALPs interacting with such a sector of dark radiation evaporate and heat up the dark radiation. This heating increases the energy density of radiation around MRE and has a potential to address the Hubble tension.
	\begin{figure*}[t]
		\centering
		\begin{minipage}{0.49\textwidth}
			\centering
			\includegraphics[width=\textwidth]{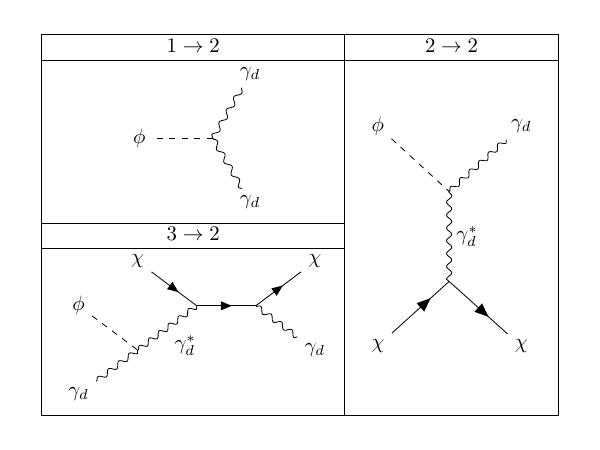}
		\end{minipage}
		%\hfill
		\begin{minipage}{0.49\textwidth}
			\centering
			\includegraphics[width=0.88\textwidth]
			{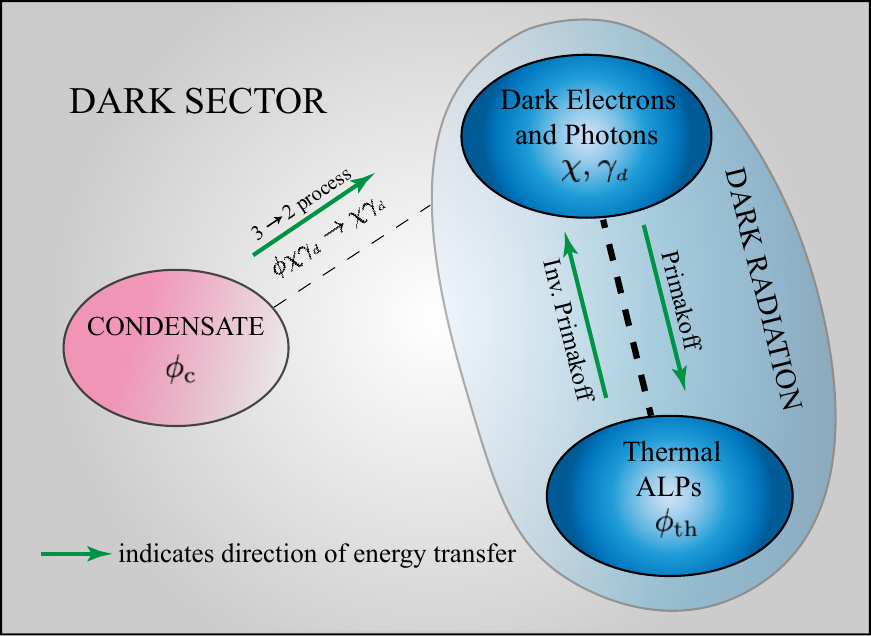}
		\end{minipage}\hfill
		\caption{\textbf{(Left)} Various processes with an ALP $\phi$ in the initial state. In the evaporation of the condensate to dark radiation, the $1\to 2$ process is switched off due to a thermal mass of the photons which is much heavier than the ALP. The $3\to 2$ process $\phi\chi\gammaD\to\chi\gammaD$, involving the exchange of an off-shell dark photon $\gammaD^*$, dominates over the $2\to2$ inverse Primakoff process, $\phi \chi \to \chi \gammaD$. \textbf{(Right)} A cartoon of the dynamics in the dark sector: the ALP condensate loses energy to the dark radiation through the $3\to2$ process, while the dark electrons and photons maintain a common temperature with the hot ALPs, through the Primakoff and inverse Primakoff process.}
		\label{fig:feyn_diag}
	\end{figure*}
	%%%%%%%%%%%%%%%%%%%%%%%%%%%%%%%
	\section{Interactions and Dark Sector Thermalization}\label{sec: dw}
	%%%%%%%%%%%%%%%%%%%%%%%%%%%%%%%
	We assume that in addition to the SM states, the universe consists of dark matter and dark radiation. We assume that the dark matter fully consists of an oscillating coherent ALP field $\phi$, with mass $m_\phi$. The dark radiation includes dark electrons $\chi$ and dark photons $\gammaD$, which are effectively massless at zero temperature and maintain a common temperature $T_\DS$. The ALP interacts with the dark photon through the usual Chern-Simons coupling, $g\phi F_d \widetilde{F}_d$, where $g \equiv \alphaD/(2\pi f)$, with $\alphaD$ being the fine structure constant of the dark sector. The dark photon interacts with the dark electron through the gauge covariant derivative. We assume that the interaction between the dark sector and the SM is negligible.
	
	We focus on the case $\mphi \ll T_\DS$, a condition required for the mechanism to work, as we will show. This scenario can occur, for instance, when the ALP is generated non-thermally, such as through the misalignment mechanism. Additionally, we require the ALP to be non-relativistic, which is ensured if the average kinetic energy of the ALP field is negligible compared to its rest mass energy. As the ALPs are the entire DM, they form a state with a high occupation number~\cite{Arias:2012az}, and effectively behave as a classical field. We will refer to ALPs in this state as the condensate in the rest of the paper.
	
	Having introduced our setup, we now discuss the dominant processes involving the ALP $\phi$. Note that $\phi$ can originate either from the condensate (which has negligible kinetic energy) or be thermal (with kinetic energy of the order of the dark radiation temperature). In our setup, we start with an initial condensate abundance, making it relevant to first consider how the condensate can lose energy, specifically by examining the dominant process involving a condensate $\phi$ in the initial state. The usually dominant $1\to2$ process, $\phi\to\gammaD\gammaD$, is suppressed for condensate $\phi$ due to the thermal mass $m_{\gammaD}^2 \sim \alphaD\,T_\DS^2$ of the dark photons when $m_\phi < m_{\gammaD}/2$. This condition is naturally satisfied since we assume $m_\phi \ll T_\DS$.
	
	The $2\to2$ process through which the condensate can lose energy is the inverse Primakoff process, $\phi\,\chi \to \gammaD\,\chi$, with a naive rate of $\Gamma_{\phi\chi \to \gammaD\chi} \propto \alphaD^4 T_\DS^3/f^2$~\cite{Braaten:1991dd, Bolz:2000fu} (see left panel of fig.~\ref{fig:feyn_diag}). However, the kinematic condition for this process is $E \gtrsim m_{\gammaD}^2/m_\phi \sim \alphaD\,T_\DS^2/m_\phi$~\cite{Arias:2012az}, where $E$ is the energy of the incoming $\chi$. For $m_\phi \ll T_\DS$, this implies $E/T_\DS \gg 1$. The abundance of electrons with such high energy is exponentially suppressed by the Boltzmann distribution, effectively making the process negligible.
	
	The dominant channel for the condensate evaporation is the $3\to 2$ process $\chi\,\phi\,\gammaD \to \chi\,\gammaD$, as shown in the left panel of fig.~\ref{fig:feyn_diag}. In the limit of small $\mphi/T_\DS$, this process can be decomposed into the ALP absorption $\phi\,\gammaD \to \gammaD^*$ followed by the Compton scattering of the off-shell photon $\gammaD^*$~\cite{Arias:2012az}. This leads to the evaporation rate for $\phi\,\chi\,\gammaD \to \chi\,\gammaD$ as:
	\begin{align}
		\Gamma_{\rm eva} \sim g^2 T^3_\DS \frac{\mphi}{\langle\Gamma_C\rangle} \sim \frac{T^2_\DS}{(2\pi f)^2} \, \mphi\,,
		\label{eq:rate}
	\end{align}
	where $\langle \Gamma_C \rangle \sim \alphaD^2\, T_\DS$ is the thermal average of the Compton scattering rate.
	\begin{figure*}[t]
		\centering
		\begin{minipage}{\textwidth}
			\centering
			\includegraphics[width=\textwidth]{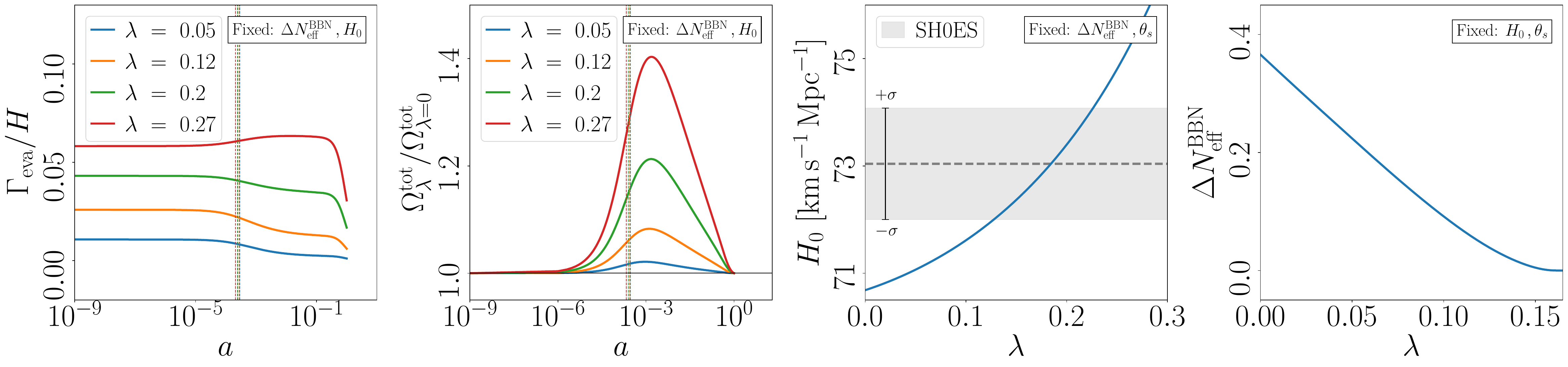}
		\end{minipage}\hfill
		\caption{The fixed parameters in each panel are taken from the best-fit point in a combined analysis of \lcdm{} + $\Delta \Neff$ to Planck 2018 and SH0ES data~\cite{Riess:2020fzl, Planck:2018vyg}, where $\Delta \NeffBBN = 0.36$, $H_0 = 70.0~\mathrm{km\, s^{-1}\, Mpc^{-1}}$, and $100\theta_s = 1.041$~\cite{Aloni:2021eaq}. \textbf{(Far Left)} Evolution of $\Gamma_{\rm eva}/H$ for cosmologies with different $\lambda$ (defined in eq.~\eqref{eq:lambda_def}) and fixed $\Delta \NeffBBN$ and \Ho{}. Note that the efficiency of evaporation in MD increases with larger $\lambda$. All interactions shut off as the universe enters dark energy domination (as seen by the sharp drop near a = 1). \textbf{(Center Left)} Evolution of total energy density in a cosmology with evaporating ALP condensate dark matter, normalized to a non-evaporating model with the same amount of extra radiation at BBN. Vertical dashed lines indicate matter-radiation equality for each model, showing minimal variation with $\lambda$ and remaining close to the standard \lcdm{} MRE. \textbf{(Center Right)} For fixed $\Delta \NeffBBN$ and $\theta_s$, $H_0$ must increase with $\lambda$ to offset changes in $r_s$. The dashed gray line shows the latest SH0ES~\cite{Riess:2021jrx} measurement of $H_0 = 73.04~\mathrm{km\, s^{-1}\, Mpc^{-1}}$, with $\pm1 \sigma = 1.04~\mathrm{km\, s^{-1}\, Mpc^{-1}}$ shaded. \textbf{(Far Right)} Degeneracy between $\Delta\NeffBBN$ and $\lambda$ for cosmologies with fixed $H_0$ and $\theta_s$. As $\lambda$ increases, less $\Delta\NeffBBN$ is required because more dark radiation is generated through evaporation.}\label{fig:solutions}
	\end{figure*}
	
	We now emphasize a crucial observation. Any process with an ALP $\phi$ in the final state, including the inverse of the processes described above, does not contribute to the energy density of the condensate. Such ALPs are produced with a thermal distribution with characteristic temperature $\TDS$. Note that for the parameter space of our interest, both the Primakoff and inverse Primakoff processes are efficient for the thermal ALP, which allows them to maintain the same temperature as the rest of the dark radiation.
	
	The ratio $\Gamma_\text{eva}/H$ behaves differently in radiation domination (RD) and matter domination (MD), scaling as $T_\DS^2/T^2$ in RD and $T_\DS^2/T^{3/2}$ in MD. For small evaporation rate, $T_\DS/T$ remains approximately constant, so that $\Gamma_\text{eva}/H$ is constant during RD and scales as $T^{1/2}$ in MD. This results in efficient evaporation during RD that slows down in MD (blue, orange and green curves in the leftmost panel of fig.~\ref{fig:solutions}). For large enough evaporation rates, $T_\DS/T$ increases over time, resulting in a different behavior of $\Gamma_\text{eva}/H$: it remains nearly constant during RD and does not decrease during MD . In this limit, the evaporation process continues and only shuts off during dark energy domination as indicated by the sharp drop near $a = 1$ for the red curve in the leftmost panel of fig.~\ref{fig:solutions}.
	%%%%%%%%%%%%%%%%%%%%%%%%%%%%%%%%%%%%%%%%%%%%%%%%%%%%%%%%%%%%
	\subsection{Boltzmann Equations}
	Given the distinct behavior of the ALP condensate compared to hot ALPs, we account for two independent contributions to the ALP energy density: one from the condensate and one from thermal ALPs. This is further motivated by the negligible interaction rate of hot ALPs with cold ALPs. As discussed earlier,  both the Primakoff and the inverse Primakoff processes are efficient for the hot ALPs, allowing them to behave like radiation. In contrast, the condensate can only lose energy through the $3\to2$ evaporation process, with the rate given in eq.~\eqref{eq:rate}. A schematic representation of the interactions between different components is shown in the right panel of fig.~\ref{fig:feyn_diag}.
	
	Let us denote the energy density of the condensate by $\rho_c$, the thermal ALP by $\rho_\phi$, and the coupled dark electron-dark photon fluid by $\rho_{\gammaD, \chi}$. The Boltzmann equations are:
	\begin{align}
		& \dot{\rho}_c + 3H \rho_c = - \Gamma_{\rm eva} \, \rho_c \,, \label{eq:boltzmann_condensate_0}\\
		& \dot{\rho}_{\gammaD, \chi} + 4H \rho_{\gammaD, \chi} = \Gamma_{\rm eva} \, \rho_c - \Gamma_{\rm prim} \, \left({\rho}_{\gammaD, \chi}- \rho_\phi \right)\, ,
		\label{eq:boltzmann_dr_0}\\%
		& \dot{\rho}_\phi + 4H \rho_\phi =  \Gamma_{\rm prim} \left({\rho}_{\gammaD, \chi}- \rho_\phi \right) \,, \label{eq:boltzmann_alp_0}
	\end{align}
	where the overdots represent time derivatives. Here $\Gamma_{\rm prim}$ schematically stands for the Primakoff and inverse-Primakoff processes that produce the hot axions. 
	It is convenient to combine eqs.~\eqref{eq:boltzmann_dr_0} and~\eqref{eq:boltzmann_alp_0} into an equation for a single energy component $\rho_{\rm DR} = \rho_\phi + \rho_{\gammaD,\chi}$.
	Moreover, since $\Gamma_{\rm prim}/H\sim (T_\DS/\mphi)\cdot (\Gamma_{\rm eva}/H)  \gg 1$, $\rho_{\rm DR}$ truly stands for the energy density of a single perfect fluid. After combining eqs.~\eqref{eq:boltzmann_dr_0}, \eqref{eq:boltzmann_alp_0} and normalizing by the constant critical density $\rho_{\rm crit} = 3H_0^2 \Mpl^2$, we obtain:
	\begin{align}
		& \dot{\Omega}_c + 3H \Omega_c = - \Gamma_{\rm eva} \, \Omega_c \,,
		\label{eq:boltz1} 
		\\
		& \dot{\Omega}_\DR + 4H \Omega_\DR = \Gamma_{\rm eva} \, \Omega_c \,,\label{eq:boltz2}
	\end{align}
	where $\Omega_i = \rho_i / \rho_{\text{crit}}$. 
	
	We define a dimensionless temperature $\widetilde{T}_\DS$, so that $\Omega_\DR = \widetilde{T}^4_\DS$. The evaporation rate and the Hubble parameter are functions of $\widetilde{T}_\DS$ and the scale factor $a$, and are given as:
	\begin{align}
		& \Gamma_{\rm eva} = \lambda\,H_0 \widetilde{T}_\DS^2  \,,\label{eq:lambda_def} \\
		& \frac{H(a)}{H_0}= \sqrt{\frac{\Omega^0_{r, \,{\rm SM}}}{a^4} + \frac{\Omega_b^0}{a^3}+\Omega_{\Lambda} + \Omega_c(a) + \widetilde{T}_\DS^4(a)}\,.
	\end{align}
	The dimensionless constant $\lambda$ determines the strength of the condensate's evaporation rate. The parameter $\Omega_{r, \,\rm{SM}}^0 \equiv \Omega^0_{\gamma} + \Omega_\nu^0$ represents today’s abundances of the SM radiation, including photons and neutrinos, $\Omega_b^0$ represents the baryon abundance today, and $\Omega_\Lambda$ is the dark energy abundance.
	
	We express the equations for $\Omega_\DR$ and $\Omega_c$ as functions of the scale factor $a$, using the relation $\mathrm{d}/\mathrm{d}t = H \, \mathrm{d}/\mathrm{d}\log{a}$. Initial conditions are set at BBN by introducing extra dark radiation, $\Omega_\DR(a_\text{BBN})$, and an initial ALP dark matter condensate, $\Omega_c(a_\text{BBN})$. The quantity $\Omega_\DR(a_\text{BBN})$ is characterized by $\Delta \NeffBBN$, measured in units of the relativistic degrees of freedom of a single neutrino. 
	
	To solve the equations and illustrate the evolution of the energy densities, we derive the dark radiation initial condition $\Omega_\DR(a_\text{BBN})$ from the \lcdm{} + $\Delta\Neff$ best-fit point from a combined analysis of Planck 2018 and SH0ES data~\cite{Riess:2020fzl, Planck:2018vyg}, where  $\Delta\NeffBBN = 0.36$~\cite{Aloni:2021eaq}. The numerical values $\Omega_{r, \rm{vis}}^0 = 8.37\times 10^{-5}$, $\Omega_b^0 = 4.55\times 10^{-2}$, and $\Omega_{\Lambda} = 0.70$ are fixed to match those of the cosmological parameters from the fit~\cite{Aloni:2021eaq}. Finally, the initial condition for the ALP condensate is determined by ensuring the solutions satisfy $\Omega^{\rm tot}(a_0) \equiv \Omega_c(a_0) + \Omega_b(a_0) + \Omega_r(a_0) + \Omega_{\Lambda} = 1$ for various values of $\lambda \neq 0$. Here, $\Omega_r$ includes contributions from SM photons, neutrinos, and dark radiation.
	
	\subsection{Evolution of Energy Densities and the Hubble Tension}
	Fig.~\ref{fig:solutions} (center left) shows the evolution of total energy density $\Omega^{\rm tot}_\lambda$ as a function of the scale factor, normalized by the total energy density with no evaporation, $\Omega^{\rm tot}_{\lambda=0}$, and with same $\Delta\NeffBBN$. As the condensate evaporates, converting energy into dark radiation, the total energy increases compared to a model with no interaction, resulting in a noticeable rise in total energy density. The higher the evaporation rate $\lambda$ is, the more pronounced is the rise. Importantly, this energy injection becomes significant only around MRE, away from BBN, acting as a natural ``on'' switch. This feature arises because the amount of dark radiation produced is proportional to the condensate energy density, which only becomes substantial near MRE. This timing is advantageous because it allows us to populate the dark radiation sector after BBN, effectively bypassing the strict BBN constraints on extra radiation~\cite{Pitrou:2018cgg, 
		Yeh:2022heq,Berlin:2019pbq, Giovanetti:2024orj}.
	
	After MRE, depending on $\lambda$, the evaporation may or may not stay efficient (see the leftmost panel of fig.~\ref{fig:solutions}). However, after MRE, the dark radiation produced by the evaporation is in MD era, and its contribution to the total energy density is subleading, getting more suppressed the farther one is from MRE. This gives a natural ``off'' switch. The ``on-off'' feature results in a bump, and is seen clearly in the center left panel of fig.~\ref{fig:solutions}.
	
	In summary, the ``on-off'' feature results from dark radiation being produced by dark matter evaporation: it is initially small during RD when dark matter energy density is subdominant, and at late times, it becomes  small again because the produced dark radiation redshifts faster than matter in MD. The effect is most pronounced during MRE, when matter and radiation are equally significant. To our knowledge, this is the first such mechanism to give an ``on-off'' feature, without a need for fine-tuning. 
	
	In the cosmologies shown in the center left panel of fig.~\ref{fig:solutions}, the values of $\Delta \NeffBBN$ and \Ho{} are kept constant. As a result, late-time parameters such as the angular diameter distance $D_A = \int_{0}^{z_{\rm rec}} \mathrm{d}z/H \propto \Ho^{-1}$ remain mostly unchanged across these cosmologies. However, early-time quantities like the comoving sound horizon $r_s = \int_{z_{\rm rec}}^{\infty} \mathrm{d}z \,(c_s/H)$ and the angular sound horizon $\theta_s = r_s/D_A$ vary, due to the added extra radiation.
	
	Given that $\theta_s$ is a precisely measured quantity, it must stay fixed. If we allow \Ho{} to vary while keeping $\theta_s$ constant across the cosmologies, we find that a non-zero $\lambda$, and consequently more dark radiation, necessitates a rise in \Ho{} to compensate for any change in $r_s$ and maintain a fixed $\theta_s$. This relationship is depicted in the center right panel of fig.~\ref{fig:solutions}, where local SH0ES measurement of the Hubble constant~\cite{Riess:2021jrx} are also overlaid. As illustrated in the plot, an evaporation rate of $\lambda \sim 0.2$ can increase \Ho{} to lie within the SH0ES band. This shows that the cosmology of evaporating ALP condensate dark matter might provide a resolution of the Hubble tension.
	
	An alternative perspective that is particularly intuitive comes from the viewpoint of constraints on extra radiation at BBN. It further highlights the possibility of evaporating ALP condensate models addressing the Hubble tension. By varying the initial radiation and examining the correlation between $\Delta \NeffBBN$ and $\lambda$, for cosmologies with fixed values of $\theta_s$ and \Ho{}, an interesting degeneracy emerges in the parameter space of $\Delta \NeffBBN$ - $\lambda$. Specifically, models with lower $\Delta \NeffBBN$ but higher $\lambda$ can achieve the same \Ho{} values as those that started with higher $\Delta \NeffBBN$ at BBN. This relationship is intuitive: even if we start with low $\Delta \NeffBBN$ values, a high evaporation rate compensates for this by the time of MRE. This approach is advantageous because it allows for an increase in \Ho{} while maintaining low levels of extra radiation near BBN, effectively bypassing the constraints imposed by BBN. This correlation is shown in the rightmost panel of fig.~\ref{fig:solutions}. 
	%%%%%%%%%%%%%%%%%%%%%%%%%%%%%
	\section{Dark Matter Abundance}\label{sec:abundance}
	%%%%%%%%%%%%%%%%%%%%%%%%%%%%%
	For the evaporation of the condensate to be efficient during RD, the evaporation rate $\Gamma_{\rm eva}$ must be of the same order as the Hubble during that period. Requiring $\Gamma_{\rm eva}/H$ to be $\mathcal{O}(1)$ leads to the relation 
	\begin{equation}\label{eq:f_m_rel}
		f^2 \sim \xi_\text{early}^2 \mphi \Mpl\:,
	\end{equation}
	where $\xi_\text{early}$ is $T_\DS/T$ well before MRE. As the condensate evaporates, $\xi$ increases from its initial value and grows further for larger values of $\lambda$. For $\lambda = 0.3$, $\xi$ evolves from an early value of $\xi_{\text{early}} = 0.045$ to approximately $\xi_{\text{late}} = 0.34$ today. For the rest of this section, we take $\xi_{\text{early}} \sim 1$ as constant; however, this does not affect the outcome of our subsequent arguments. Eq.~\eqref{eq:f_m_rel} relates $\mphi$ to $f$, just as the relation $m_\phi = m_{\pi}f_{\pi}/f$ does for the QCD axion.
	
	For an ALP condensate produced via the standard misalignment mechanism to account for all of today's dark matter, we have~\cite{Marsh:2010wq,Blinov:2019rhb}
	\begin{align}
		1 = \frac{\Omega_c}{\Omega_{\rm DM}} \simeq \frac{1}{1.2 \times 10^{19}}\left(\frac{f \theta_i}{\text{GeV}}\right)^2 \left(\frac{\mphi}{\text{GeV}}\right)^{1/2}\,.
		\label{eq:abund}
	\end{align}
	Assuming $\theta_i$ to be $\mathcal{O}(1)$ and using eq.~\eqref{eq:f_m_rel}, we get
	\begin{equation}
		m_a \sim 2.9~\text{GeV}, \quad f \sim 2.6 \times 10^9~\text{GeV}\,.
	\end{equation}
	
	However, this is problematic because to prevent the ALP condensate from decaying into dark radiation at any point in the cosmological history, we need $\mphi \ll T_\DS^{\rm today} \sim T^{\rm today} \sim \text{meV}$. To resolve this, we need production mechanisms that can achieve the necessary condensate abundance for it to constitute $100\%$ of dark matter while keeping the ALP mass small. We consider two such mechanisms: $i)$ the clockwork mechanism~\cite{Choi:2014rja,Choi:2015fiu, Kaplan:2015fuy} and $ii)$ the kinetic misalignment mechanism (each are discussed in more detail in Appendices~\ref{app:clockwork} and~\ref{app:kin-mis}).
	
	Among the two, we find that the clockwork mechanism can resolve the issue by introducing up to $10$ additional heavy ALP fields in the ultraviolet (UV). For kinetic misalignment, we find that the necessary condition from~\cite{Co:2019jts} for maximum enhancement are met in our condensate mode. However, to satisfy the sufficient conditions, we would need to specify a UV completion of our model. We leave this exploration for future work.
	
	\section{Outlook}
	In this letter, we have investigated a new phenomenon that occurs when non-thermally produced ALP dark matter interacts with a sector of dark radiation. We demonstrated that, in the presence of dark radiation, ALPs evaporate into dark radiation, increasing the energy density around the time of MRE. This ALP evaporation process could have measurable effects on cosmological observables. Specifically, we explored how this phenomenon could influence the expansion rate of the universe and highlighted its potential to address the Hubble tension. 
	
	It remains to be seen whether this mechanism will provide a viable solution to the Hubble tension. A thorough analysis using Markov Chain Monte Carlo (MCMC) methods can be employed to explore the parameter space and compare the model with real CMB data, such as that measured by Planck.
	
	Our preliminary study raises several open questions that require further investigation. One key issue concerns the production mechanism of ALPs. We have shown that the standard misalignment mechanism is not viable in this context and have identified alternative known mechanisms that can be effective. However, a thorough exploration is needed to fully understand the UV completion of these production mechanisms. In that context, it also remains to be seen if a similar mechanism can work with other light fields.
	
	Another question involves additional astrophysical and cosmological observables. The evaporation of dark matter could significantly impact structure formation, the growth rate of cosmic structures, and even the evolution of galaxies. Two competing effects come into play: (i) the presence of more matter in the early universe accelerates structure growth compared to $\Lambda$CDM, and (ii) since the ALP evaporation rate is proportional to the ALP density, denser regions evaporate more, resulting in a more uniform universe compared to $\Lambda$CDM. It is important to note that the evaporation rate, $\Gamma_{\rm eva}/H$, only decreases rapidly when dark energy starts to dominate the universe's energy density. While this wasn't crucial for understanding the universe's expansion rate, it becomes essential when studying structure formation. A dedicated analysis is needed to determine which of these competing processes has a more significant influence.
	
	Lastly, exploring axion evaporation in scenarios involving interactions with SM photons presents an intriguing opportunity. Such interactions could induce spectral distortions in the CMB blackbody spectrum, offering new observational signatures. This approach could provide a novel way to test the interplay between dark matter and visible sector particles, deepening our understanding of axions and their role in the universe.
	
	\section*{Acknowledgements}
	We thank Matt Reece and Martin Schmaltz for useful discussions and comments on the manuscript. HB thanks David J. E. Marsh for helpful discussion on ALP abundance. RKM thanks discussions with Marco Raveri. The work of DA is supported by the U.S. Department of Energy (DOE) under Award DE-SC0015845. HB is supported by the DOE Grant DE-SC0013607. The work of RKM is supported by NSF grants PHY-1620806, PHY-1748958 and PHY-1915071, the Chau Foundation HS Chau postdoc award, the Kavli
	Foundation grant ``Kavli Dream Team'', and the Moore Foundation Award 8342. 
	
	\bibliography{AE_bib.bib}
	
	\clearpage
	\onecolumngrid
	\appendix
	
	%%%%%%%%%%%%%%%%%%%%%%%%%%%%%
	% APPENDICES
	%%%%%%%%%%%%%%%%%%%%%%%%%%%%%
	
	%%%%%%%%%%%%%%%%%%%%%%%%%%%%%
	\section{Dark Matter Abundance from Clockwork}
	\label{app:clockwork}
	%%%%%%%%%%%%%%%%%%%%%%%%%%%%%
	\begin{figure}[t]
		\centering
		\begin{minipage}{\textwidth}
			\centering
			\includegraphics[width=\textwidth]{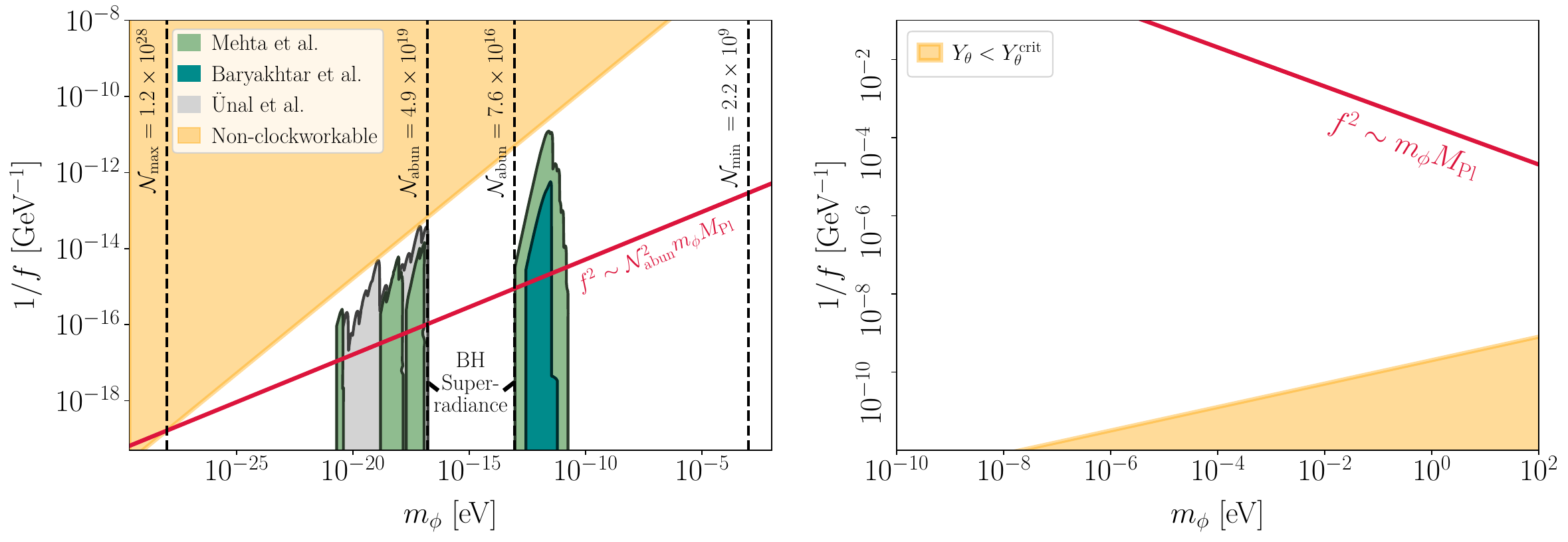}%
	\end{minipage}\hfill
	\caption{\textbf{(Left)} The constraints on the ALP mass and decay constant are shown. ALPs along the crimson line can potentially address the Hubble tension and give the correct dark matter abundance today through the clockwork mechanism. Vertical dashed black lines indicate the needed value of $\mathcal{N}_{\rm abun}$ for a given $m_\phi$, using eq.~\eqref{eq:N_abun}. The black hole  superradiance bounds are from~\cite{Mehta:2021pwf, Baryakhtar:2020gao,Unal:2023yxt, Unal:2020jiy}. The plots are generated from the public code~\cite{AxionLimits}. \textbf{(Right)} The allowed parameter space for ALPs generated through the kinetic misalignment mechanism: the white region satisfies the necessary condition for the mechanism to work and for the ALPs to account for all of dark matter today. ALPs along the crimson line can potentially address the Hubble tension.
	}
	\label{fig:abundance} 
\end{figure}

The clockwork mechanism~\cite{Choi:2014rja,Choi:2015fiu, Kaplan:2015fuy} generates an ALP with a significant difference between the effective decay constant $f$ in the ALP potential and the fundamental decay constant $F$ in the Chern-Simons (CS) coupling. Let's  focus on the ALP potential and the CS coupling terms 
\begin{align}
	\mathcal{L} \supset \Lambda^4 \cos{\frac{\phi}{f}} + k\frac{\alphaD}{8\pi} \frac{\phi}{F} F_{\mu\nu} \widetilde{F}^{\mu\nu} = \Lambda^4 \cos{\frac{\phi}{f}} + jk\frac{\alphaD}{8\pi} \frac{\phi}{f} F_{\mu\nu} \widetilde{F}^{\mu\nu}\:,
	\label{eq:lagrangian}
\end{align}
where $k$ is an integer, and we have written $f = j F$ in the second equality. The cosine potential for the ALP arises from the non-perturbative dynamics of a confining gauge group with confinement scale $\Lambda$. The integer $k$ is a group-theoretic factor whose specific value depends on the details of the UV completion, e.g. in KSVZ UV completion, $k$ is the Dynkin index of the representation of heavy fermions under the gauge group. The coefficient $j$ must be an integer to preserve the ALP shift symmetry and can come from monodromy~\cite{Kim:2004rp} or by integrating out a tower of confining gauge groups~\cite{Agrawal:2017cmd}. 

As eq.~\eqref{eq:lagrangian} shows, the ALP coupling to $F\widetilde{F}$ is enhanced by a factor $\mathcal{N} \equiv jk$, which separates the fundamental and effective ALP decay constants. The condition between $f$ and $m_\phi$, previously derived in eq.~\eqref{eq:f_m_rel} by requiring $\Gamma_\text{eva}/H\sim\mathcal{O}(1)$ during RD, is modified to (taking $\xi\sim1$)
\begin{align}
	f^2\sim \mathcal{N}^2\, \mphi\Mpl\,.
	\label{eq:f_m_rel_N}
\end{align}
Plugging this new relation between $f$ and $m_\phi$ into eq.~\eqref{eq:abund}, which ensures that the ALPs account for all of the dark matter, we get
\begin{align}
	\mathcal{N}_{\rm abun}\sim \left( \frac{2.9~\rm{GeV}}{\mphi}\right)^{3/4}\,,
	\label{eq:N_abun}
\end{align}
where $\mathcal{N}_{\rm abun}$ is the exact amount of enhancement needed to achieve the correct dark matter abundance for any given ALP mass. Requiring the ALP mass to remain below $\rm{meV}$ imposes a lower bound on the enhancement factor:
\begin{align}
	\mathcal{N}_{\rm abun} \gtrsim \mathcal{N}_{\rm min}\sim 2.2 \times 10^9\,. 
\end{align}

Even though the clockwork mechanism seems to solve the abundance issue, caution is needed when dealing with large enhancement factors. There is an upper bound on how large the clockwork enhancement can be~\cite{Agrawal:2017cmd, Bagherian:2022mau}. This limit arises due to individual upper bounds on integers $k$ and $j$, which we explain below.
\begin{itemize}
	\item The integer $k$ is the Dynkin index of the representation of heavy fermions in a KSVZ setup. Above the confinement scale, in the non-canonical normalization, the gauge field propagator receives fermionic loop corrections proportional to $\mathrm{Tr}[\tau^a \tau^b] \propto k$, where $\tau^i$ are the gauge group generators. To keep these corrections subdominant relative to the tree-level gauge field propagator $\propto \alphaD^{-1}$ and maintain perturbativity we require $k \lesssim \alphaD^{-1}$.
	\item The integer $j$ arises as a product of smaller integers, each from integrating out a heavy ALP in an \textit{axion alignment} setup. Consider two ALPs $\phi_1, \phi_2$ with decay constants $f_1, f_2$ respectively, and the potential 
	\begin{align}
		V(\phi_1, \phi_2) \supset \Lambda_1^4 \cos{\frac{\phi_1}{f_1}} + \Lambda_2^4 \cos{\left(\frac{\mathfrak{j} \phi_1}{f_1} + \frac{\phi_2}{f_2}\right)},
	\end{align}
	where $\mathfrak{j}$ is an $\mathcal{O}(1)$ integer. When $\Lambda_2 \gg \Lambda_1$, one linear combination is heavy compared to the other and can be integrated out. This forces the lighter mode to traverse a longer path in field space, resulting in an effective potential for $\phi_2$ (the mostly lighter axion) with decay constant $f = \mathfrak{j} f_2$:
	\begin{equation}
		V_{\rm eff.}(\phi_2) \supset \Lambda_1^4 \cos{\frac{\phi_2}{\mathfrak{j} f_2}}.
	\end{equation}
	
	Extending this mechanism to an \textit{n-site alignment} yields a large enhancement, $j = \prod_i \mathfrak{j}_i$. In this scenario, each cosine term mediates ALP scattering and must obey perturbative unitarity bounds, leading to $j < \frac{f}{2\pi \Lambda}$~\cite{Agrawal:2018mkd}.
\end{itemize}

Combining both upper bounds gives the clockwork constraint on the ALP coupling to dark photons:
\begin{align}
	\mathcal{N}\leq \frac{f}{2\pi\alphaD\,\Lambda} = (2\pi\alphaD)^{-1}\sqrt{f/\mphi}\,,
	\label{eq:cw_const}
\end{align}
where in the last term we have use $\mphi^2 = V''(\phi/f=\pi) = \Lambda^4/f^2$. 

The inequality~\eqref{eq:cw_const} restricts the parameter space in the $\mphi$ - $f$ plane, as shown by the shaded orange region in the left panel of fig.~\ref{fig:abundance}. The equation for this region is obtained by substituting $\mathcal{N}_{\rm abun}$ from eq.~\eqref{eq:N_abun} into~\eqref{eq:cw_const}. We have taken $\alphaD \sim 0.1$ here. The crimson line in the left panel of fig.~\ref{fig:abundance} corresponds to ALP condensate satisfying eq.~\eqref{eq:f_m_rel_N} as well as providing the correct dark matter abundance. We observe that there is significant parameter space where all the conditions are met. Using Stirling's approximation and assuming the primary enhancement arises from the $n$-site alignment, we estimate that at least $j \sim 10$ heavy axions must be integrated out to reach the allowed parameter space (since $N_\text{min} \sim 10^{10}$).

The clockwork mechanism fails to produce the correct dark matter abundance for arbitrarily light ALPs. As shown, $\mathcal{N}_{\rm abun}$ reaches the clockwork upper bound around $\mphi \sim 10^{-28}~\rm{eV}$, corresponding to $\mathcal{N}_{\rm max} \sim 1.2 \times 10^{28}$. However, this is of little concern since ALPs with masses below $10^{-19}~\rm{eV}$ face constraints from large-scale structure observations such as Lyman-$\alpha$, weak lensing, and even CMB Planck data~\cite{Antypas:2022asj,Feix:2019lpo,Winch:2024mrt,Rogers:2023ezo,Rogers:2020ltq}. It would be interesting to consider the ALP in the region between bounds from black hole superradiance~\cite{Mehta:2021pwf, Baryakhtar:2020gao,Unal:2023yxt, Unal:2020jiy}. Analyzing the phenomenological consequences of this bound is left for future work.

%%%%%%%%%%%%%%%%%%%%%%%%%%%%%
\section{Dark Matter Abundance from Kinetic Misalignment}
\label{app:kin-mis}
%%%%%%%%%%%%%%%%%%%%%%%%%%%%%

In previous sections, we assumed a temperature-independent ALP mass, the simplest scenario. Here, we consider a temperature-dependent ALP mass inspired by the QCD axion mass-temperature relation~\cite{Blinov:2019rhb}. Similar to the QCD axion, we assume that when the temperature of the dark sector is below the confinement scale $\Lambda$, the mass is temperature independent, and has a specific dependence for higher temperatures. We take the temperature dependence of the ALP mass to be
\begin{align}
	\mphi(T) = 
	\begin{cases}
		\mphi^{\rm IR} & \text{if } \TDS < \Lambda\\
		\mphi^{\rm IR}\left(\frac{\TDS}{\Lambda}\right)^{-4} & \text{if } \TDS \geq \Lambda
	\end{cases}\qquad ,
\end{align}
where $m_\phi^\text{IR} = m_\phi(0)$ is the zero temperature mass of the ALP and  $\Lambda^2 \equiv \mphi f$. To prevent extra relativistic degrees of freedom in the UV theory from affecting $\Delta \Neff$ bounds from the CMB, we require $T_{\Lambda} > T_{\rm BBN} \sim 5~\rm{MeV}$, where $T_{\Lambda} = \Lambda/\xi$ is the SM temperature corresponding to the dark confinement scale $\Lambda$.

In the standard misalignment mechanism, the ALP begins oscillations at the standard model temperature $T = \Tosc$ when $m(\Tosc) = q H(\Tosc)$, where $q$ is a numerically derived proportionality constant. As noted in ref.~\cite{Blinov:2019rhb} $q = q_0\sim 1.6$ when $\Tosc{} < T_\Lambda = \Lambda/\xi$ and $q = q_T \sim 4.8$ when $\Tosc{} > T_\Lambda = \Lambda/\xi$. Our second model-building technique that addresses the dark matter abundance issue involves a non-standard misalignment scenario known as kinetic misalignment~\cite{Co:2019jts}. Unlike the standard mechanism, where the ALP starts from stationary initial conditions, in this scenario, the ALP begins with significant initial kinetic energy i.e. $\dot{\theta}_\text{initial} \equiv \dot{\phi}_\text{initial}/f \neq 0$. Discussions in this section closely follow the work of~\cite{Co:2019jts}, where further details are provided.

If the initial kinetic energy, $K = \dot{\theta}^2 f^2 / 2$, in the kinetic misalignment scenario exceeds the maximum potential $V_{\max}(\Tosc) \equiv 2\mphi^2(\Tosc)f^2$ at the standard $\Tosc$, the ALP overcomes the barrier and continue rolling. The condition for kinetic misalignment is therefore:

\begin{equation}\label{eq:mis_cond1}
	\dot\theta^2(\Tosc)f^2/2 > 2\mphi^2(\Tosc)f^2\,.
\end{equation}

The value of $\dot\theta(\Tosc)f^2$ depends on the UV completion of the theory. Ref.~\cite{Co:2019jts} discusses cases involving quadratic and quartic potentials for the global symmetry breaking field. Care must be taken to ensure that the decay of the radial mode in the UV completion into dark photons does not alter the Boltzmann equations or the dark radiation energy density. In this paper, we assume a UV completion that does not affect the radiation density, and satisfies~\eqref{eq:mis_cond1}, leaving the detailed model construction and constraints for future work.

To parametrize the kinetic energy of the ALP, ref.~\cite{Co:2019jts} defines the \textit{yield} $Y_{\theta}$ as
\begin{align}
	Y_{\theta}(T) \equiv \frac{\dot{\theta}(T)f^2 }{s(T)} \approx \text{constant},
\end{align}
where $\dot{\theta}f^2 $ is the Noether charge associated with the ALP shift symmetry $\phi \rightarrow \phi+ \alpha f$, and scales as $a^{-3} \propto T^3$, with $a$ being the cosmological scale factor. Since the entropy density $s(T) \propto T^3$, $Y_{\theta}(T)$ remains approximately constant. We can rewrite the condition for kinetic misalignment, i.e., eq.~\eqref{eq:mis_cond1}, in terms of \( Y_{\theta} \) by normalizing it with \( s(\Tosc)/f^2 \):
\begin{align}
	Y_{\theta}=\frac{\dot{\theta}(\Tosc)f^2}{s(\Tosc)} > \frac{2\mphi(\Tosc)f^2}{s(\Tosc)}\sim \frac{2q_TH(\Tosc)f^2}{s(\Tosc)}\sim Y_{\rm crit}\equiv \frac{2q_Tf^2}{\Mpl\Tosc}\,.
	\label{eq:mis_cond3}
\end{align}
In the final step, we used $H/s \propto 1/T$ and the relation between the Hubble parameter and the potential at $\Tosc$. Thus, kinetic misalignment occurs when $Y_{\theta} > Y_{\rm crit}$.

In this scenario as the ALP continues rolling, the misalignment angle evolves at a rate $\dot{\theta}$ until a new oscillation temperature \Tosck{} is reached, where the redshifted kinetic energy balances the maximum potential energy,

\begin{equation}\label{eq:mis_cond2}
	\dot{\theta}(\Tosck) = 2\mphi(\Tosck)\,.
\end{equation}
If this occurs after the traditional $\Tosc\geq \Tosck$, then kinetic misalignment can enhance the dark matter abundance, as we discuss next.

The ratio of ALP number density to entropy density, $n_{\phi}/s$, is also a redshift independent quantity after oscillations begin at $\Tosck$. Thus one can relate $Y_{\theta}$ to the ALP abundance value right after oscillations begin: 
\begin{equation}
	\frac{\rho_\phi}{s} =\mphi^{\rm IR}\, \frac{n_\phi(\Tosck)}{s(\Tosck)} = \mphi^{\rm IR}\, \frac{V_{\phi}(\Tosck)/\mphi^{\rm IR}}{s(\Tosck)}\sim \mphi^{\rm IR}\, \frac{2\mphi(\Tosck)f^2}{s(\Tosck)}\sim \mphi^{\rm IR}\, \frac{\dot{\theta}(\Tosck)f^2}{s(\Tosck)} = \mphi^{\rm IR}\, Y_{\theta}\,,
\end{equation}
where the final step uses~\eqref{eq:mis_cond2}. This parametrization allows us to express the ratio of ALP relic density to dark matter density as
\begin{equation}
	\frac{\Omega_{\phi}}{\Omega_{\rm DM}} = \frac{\rho_{\phi}}{\rho_{\rm DM}} = \frac{\mphi^{\rm IR}\, Y_{\theta}}{0.44\,\mathrm{eV}}\,.
\end{equation}
If ALPs constitute all dark matter, we need $Y_{\theta}= 0.44\,\mathrm{eV} /\mphi^{\rm IR}$. Requiring $Y_{\theta} > Y_{\rm crit}$ provides the condition for ALPs to account for all dark matter via kinetic misalignment:
\begin{equation}
	\frac{2q_Tf^2}{\Mpl \Tosc} < \frac{0.44\, \rm{eV}}{\mphi^{\rm IR}}\,.
\end{equation}
Using $m(\Tosc) = q_T H(\Tosc)$ and assuming $m(\Tosc) = \mphi^{\rm IR} \left(\frac{\xi \Tosc}{\Lambda}\right)^{-4}$ before confinement, we find
\begin{align}
	\left(\frac{f}{\Mpl}\right)^5 
	< 
	\frac{1}{8\xi^2}
	q_T^{-7/2}
	\left(\frac{0.44 \text{eV}}{m_\phi^\text{IR}}\right)^{3/2}
	\left(\frac{0.44 \text{eV}}{M_\text{pl}}\right)^{3/2}
\end{align}
The orange-shaded regions in right panel of fig.~\ref{fig:abundance} are excluded by this constraint. The crimson line corresponding to the condensate potentially resolving the Hubble tension is also shown. As illustrated, these ALPs are not excluded by the conditions for kinetic misalignment. However, not all of the white region can be utilized, as specific UV completions impose further constraints, often coming from the decay of the heavier modes (e.g. see discussions in ref.~\cite{Co:2019jts}).
\end{document}